# Fundamental theorems of traffic flow

Xin Shui*

*School of Traffic Management, People's Public Security University of China, Beijing, China*
*\* Corresponding author: shuixin@ppsuc.edu.cn (X. Shui).*

## ABSTRACT

To gain essential understandings of traffic flow, four theorems are derived to establish the kinematics of the basic unit of traffic flow, namely two consecutive vehicles. The first is to determine the two critical distances of the vehicle spacing, the second is to derive the spacing margin and the critical distance difference as well as the relationship they must satisfy, the third is to complete the mathematical definition and give the necessary and sufficient condition for the time headway to remain constant, and the fourth is to derive the governing equation of the spacing margin and determine the two key constraint spaces of the motion of consecutive vehicles. On this basis, the composition of the traffic flow phenomenon and that of the traffic flow property are determined, and the inherent regularity of traffic flow is briefly elucidated. Thus, a basic theoretical framework specific to the study of traffic flow is obtained. Based on the framework, three important issues, namely traffic flow instability, phase, and empirical datasets, are briefly reviewed.

## 1. Introduction

Traffic flow has been studied for approximately 90 years [1-3], and a tremendous amount of work has been proposed. However, there are still many problems in the theory of traffic flow that have not been discovered or solved.

First, the scarcity of empirical data leads to the insufficient discoveries and unclear classifications of traffic flow phenomena. For instance, Kerner proposed the three-phase traffic theory to qualitatively describe the spatiotemporal features of congested traffic patterns [4-6], but this theory has triggered a long debate [7, 8], and no final conclusion has been reached yet [9].

Second, the incomprehensive considerations of traffic flow phenomena result in the incomplete definitions of some basic concepts. The current mathematical definition of the time headway does not explicitly take into account the situation where the vehicle speed is zero at a measurement location [10-12], and the current definition of the reaction time does not take into account the driving operation and the motion state of the vehicle [13-15].

Third, the inadequate understandings of traffic flow properties make it difficult to reveal causes and regularities of phenomena such as disturbance growth. In previous studies, many factors have been shown to affect the stability of traffic flow [16-25], but no study to date has been able to synthesize these findings, and the complete evolution of the disturbance is still not well revealed.

Fourth, the lack of a theoretical study on the motion of consecutive vehicles is a major impediment to gaining essential understandings of traffic flow. Despite hundreds of models having been proposed, modeling is still the core of the traffic flow study. The basis of the modeling is to empirically assume, simplify, or even ignore (employ some other physical theories) the complex motion of vehicles [26-40], which, while approximating or explaining the traffic phenomenon to some extent, fails to lead to the essence.

Researchers tend to be convinced that "traffic theory is inherently an experimental science and should be pursued as such" [41], and that traffic flow studies should be based on large empirical datasets. It is worth noting that observation and experimentation are always the basis for the development of natural sciences, and the experimental science or the theoretical science is more of the category distinguished by the methodology. Newton's classical mechanics is a model of experimental sciences, but it is also a theoretical system based on models, laws, theorems, etc. In fact, the methodology of the experimental science should ultimately verify principles or reveal laws, which cannot be achieved by logical proofs, or should lead to and corroborate theorems derived from reasoning, thereby forming a theoretical framework, while other discoveries are in the intermediate or applied stage. Traffic flow, although it is a physical system integrated with human drivers or intelligent driving systems, is always governed by the laws of kinematics and dynamics. Whether the traffic theory has its own principles or laws is still unclear, but it is certain that based on existing empirical studies, the fundamental theorems of traffic flow can be derived by applying mechanics, mathematics, etc. These theorems will establish the kinematics of two consecutive vehicles, which is undoubtedly conducive to the further development of researches.

The rest of the paper is organized as follows. In Section 2, four fundamental theorems are derived to establish the kinematics of two consecutive vehicles. On this basis, Section 3 determines the composition of the traffic flow phenomenon and that of the traffic flow property, and briefly elucidates the inherent regularity of traffic flow. Thus, a basic theoretical framework specific to the traffic flow study is obtained. Based on the framework, Section 4 briefly reviews three important issues in traffic flow study, namely traffic flow instability, phase, and empirical datasets, as the conclusion of this paper.

## 2. Four fundamental theorems

### 2.1. Theorem 1: Two critical distances

In traffic flow, the spacing between two consecutive vehicles is a variable with two critical distances, namely, *a priori* safety critical distance (APSCD) and *a posteriori* collision critical distance (APCCD).

At time $t$, the preceding vehicle (PV, subscript $n-1$, see Fig. 1) and the following vehicle (FV, subscript $n$) move with the speeds $v_{n-1}(t)$ and $v_n(t)$, and the accelerations $a_{n-1}(t)$ and $a_n(t)$, respectively. If the PV starts braking with its minimum deceleration (variable, depending on the maximum braking moment, the air resistance, the friction, etc.) at $t$ until it stops, in order to avoid collisions (without changing lanes), the FV also needs to start braking (usually after some time) until it stops. According to kinematics, there is a minimum safety spacing between the PV and the FV at $t$, namely

$$D_n^S(t) = \max\left\{\int_t^u \left(v_n^{d_{\min}}(\tau) - v_{n-1}^{d_{\min}}(\tau)\right)d\tau\right\}, u \in I_n^{\min}(t), \tag{1}$$

where $I_n^{\min}(t) = \left[t,\ t + \tau_n^{\min}(t) + d_{v_n \to 0}^{\min}\left(t + \tau_n^{\min}(t)\right)\right]$, and $\tau_n^{\min}(t)$ and $d_{v_n \to 0}^{\min}\left(t + \tau_n^{\min}(t)\right)$ are the minimum values of the reaction time (from $t$) and braking-to-stop duration (from $t + \tau_n^{\min}(t)$) of the FV; $v_n^{d_{\min}}$ and $v_{n-1}^{d_{\min}}$ are the speeds (may not actually happen) when the motion durations of the FV and the PV are minimum, and $v_n^{d_{\min}}(t) = v_n(t)$ and $v_{n-1}^{d_{\min}}(t) = v_{n-1}(t)$. The motion process of the FV over the interval $I_n^{\min}(t)$ is called the emergency reaction-braking process (ERBP).

$D_n^S$ is the APSCD, which is *a priori* because when vehicle spacing $S_n(t) \geq D_n^S(t)$, the FV will never collide with the PV as long as the ERBP is started, regardless of the motion of the PV. That is, the vehicle spacing greater than or equal to $D_n^S$ can ensure the safety of the car-following motion. Therefore, the APSCD serves as the first critical distance.

However, it should be noted that since the PV does not always brake to a stop with its minimum deceleration, the FV does not necessarily collide with the PV when $S_n(t) < D_n^S(t)$. The inevitable collision means that even if the FV starts the ERBP at *t*, there is still not enough spacing to avoid a collision with the PV. Thus, the APCCD at *t* is

$$D_n^C(t) = \max\left\{\int_t^u \left(v_n^{d_{\min}}(\tau) - v_{n-1}(\tau)\right)d\tau\right\}, \ u \in I_n^{\min}(t), \tag{2}$$

where $v_{n-1}$ is the real speed of the PV. According to kinematics, when $S_n(t) < D_n^C(t)$, the PV and the FV will have an inevitable collision if there is no lane change. Hence, the APCCD serves as the second critical distance of the vehicle spacing. It is *a posteriori* because, for the FV, the speed change of the PV cannot be accurately predicted in an environment where there is no autonomous driving and no vehicle-to-vehicle or vehicle-to-infrastructure communication.

## 2.2. Theorem 2: Spacing margin and critical distance difference

In traffic flow, the spacing between two consecutive vehicles can be divided into two parts: the APSCD and the spacing margin. The variation of the vehicle spacing is the variation of these two parts, where the spacing margin can be negative. The difference between the APSCD and the APCCD is the critical distance difference (CDD), and the sum of the CDD and the spacing margin must be greater than or equal to zero when no lane change occurs.

Let $M_n$ denote the spacing margin. Then the spacing between the PV and the FV

at time $t$ can be written as

$$S_n(t) = D_n^S(t) + M_n(t). \tag{3}$$

If the time changes by an increment $\Delta t$, then the corresponding increment of the spacing is

$$S_n(t+\Delta t) - S_n(t) = \int_t^{t+\Delta t} (v_{n-1}(\tau) - v_n(\tau)) d\tau. \tag{4}$$

Next, the increment of the spacing margin can be derived as

$$M_n(t+\Delta t) - M_n(t) = \int_t^{t+\Delta t} (v_{n-1}(\tau) - v_n(\tau)) d\tau - (D_n^S(t+\Delta t) - D_n^S(t)). \tag{5}$$

If $M_n(t)$ is differentiable at $t$, then

$$\frac{dM_n(t)}{dt} = v_{n-1}(t) - v_n(t) - \frac{dD_n^S(t)}{dt}. \tag{6}$$

Let $D_n^D$ denote the CDD, then

$$D_n^D(t) = D_n^S(t) - D_n^C(t) \tag{7}$$

and the following inequality should be satisfied:

$$M_n(t) + D_n^D(t) \geq 0. \tag{8}$$

*2.3. Theorem 3: Constant time headway*

In traffic flow, two consecutive vehicles keep a constant time headway if and only if the FV replicates the spatiotemporal trajectory of the PV, that is, having the same speed, acceleration, and jerk at the same location. The time headway changes depending on the speed difference between the two consecutive vehicles passing through the same location at different times.

Time headway is defined as the time interval during which the front edges of two consecutive vehicles pass through the same location. The location where the front edge of the PV is located at a certain time is usually taken as a measurement location, namely $x_{n-1}(t)$. And the time at which the speed of the PV (or the FV) is positive at the measurement location is the start time (or the end time) of the time headway measurement, denoted by $t_{n-1}^s(x_{n-1}(t),\ldots)$ (or $t_n^e(x_{n-1}(t),\ldots)$), then

$$t_{n-1}^s(x_{n-1}(t),\ldots) = t + \int_t^{+\infty} \text{sgn}\left(1 - \text{sgn}\left(\frac{x_{n-1}(\tau) - x_{n-1}(t)}{d\tau}\right)\right) d\tau, \tag{9}$$

$$t_n^e(x_{n-1}(t),\ldots) = t_{n-1}^s(x_{n-1}(t),\ldots) + h_n(x_{n-1}(t),\ldots), \tag{10}$$

where $h_n(x_{n-1}(t),\ldots)$ is the time headway between the PV and the FV, and

$$h_n(x_{n-1}(t),\ldots) = \int_{t_{n-1}^s(x_{n-1}(t),\ldots)}^{+\infty} \text{sgn}\left(1 - \text{sgn}\left(\frac{x_n(\tau) - x_{n-1}(t)}{d\tau}\right)\right) d\tau, \tag{11}$$

where $x_n$ is the location of the front edge of the FV. Here the forward direction of the two vehicles is the default positive direction. It is clear that $h_n$ remains constant for any increments of $t$ if and only if the FV replicates the spatiotemporal trajectory of the PV, and if the speed of the FV passing through each location is greater (or less) than that of the PV passing through that location, $h_n$ is decreasing (or increasing) over time.

*2.4. Theorem 4: Two key constraint spaces*

In traffic flow, when the FV passes through the location where the PV has passed, the feasible speed and acceleration depend on the time headway between the two vehicles, the motion of the PV and its bodywork length, and the ERBP of the FV, etc., and two key constraint spaces exist.

When the front edge of the FV passes through the location $x_{n-1}(t)$, the spacing margin is governed by

$$M_n(t_n^e) = \int_{t_{n-1}^s}^{t_n^e} v_{n-1}(\tau) d\tau - L_{n-1} - D_n^S(t_n^e), \tag{12}$$

where $L_{n-1}$ is the bodywork length of the PV. Then $t_n^e$ and the corresponding speed and acceleration of the FV must belong to the following set:

$$C_n^s = \left\{(t_n^e, v_n(t_n^e), a_n(t_n^e)) \mid (t_n^e, v_n(t_n^e), a_n(t_n^e)) \in C_n \wedge M_n(t_n^e) + D_n^P(t_n^e) \geq 0\right\}, \tag{13}$$

where

$$C_n = \left\{(t_n^e, v_n(t_n^e), a_n(t_n^e)) \middle| \begin{array}{l} \int_{t_{n-1}^s}^{t_n^e} j_n(\tau) d\tau = a_n(t_n^e) - a_n(t_{n-1}^s), \\ \int_{t_{n-1}^s}^{t_n^e} a_n(\tau) d\tau = v_n(t_n^e) - v_n(t_{n-1}^s), \\ \int_{t_{n-1}^s}^{t_n^e} v_n(\tau) d\tau = x_{n-1}(t) - x_n(t_{n-1}^s), \\ 0 \leq v_n(\cdot) \leq v_n^{\lim}(\cdot) \end{array}\right\}, \tag{14}$$

and $j_n$ is the jerk (determines the range of $a_n$) and $v_n^{\lim}$ is the speed limit. This is a strict constraint space. If the motion of the PV cannot be accurately predicted, the set should be

$$C_n^{ls} = \left\{ \left( t_n^e, v_n(t_n^e), a_n(t_n^e) \right) \middle| \left( t_n^e, v_n(t_n^e), a_n(t_n^e) \right) \in C_n \wedge M_n(t_n^e) \geq 0 \right\}, \tag{15}$$

which constitutes a less stringent constraint space.

## 3. Essential understandings of traffic flow

*3.1. Composition of the traffic flow phenomenon*

According to Theorems 1 and 2, $0 \leq D_n^C(t) \leq D_n^S(t) \leq \int_{I_n^{\min}(t)} v_n^{d_{\min}}(\tau) d\tau$ is always true, and the change of the spacing margin between two consecutive vehicles can be divided into 3 stages, denoted by SMS. 1 to 3 in Table 1.

According to Theorems 1 to 3, the time headway, spacing margin, and CDD change depending on the relationship between the current motion of the FV and the spatiotemporal trajectory that the PV has traveled, is traveling, and is and will be traveling, respectively. Therefore, there are 27 motion stages for two consecutive vehicles, denoted by MS. 1 to 27 in Table 2.

The traffic flow phenomenon is the motion of consecutive vehicles. Therefore, the composition of the traffic flow phenomenon comprises the motion stages of consecutive vehicles together with the spacing margin stages (SMSs). Different traffic flow phenomena correspond to different stabilities, transition sequences, and occurrence frequencies of the motion stages, as well as the different spacing margin stages.

*3.2. Composition of the traffic flow property: commonality and heterogeneity*

In traffic flow, the speed of the PV and the spacing and relative speed (i.e., the rate of change of the spacing) between the two consecutive vehicles constitute the composition of the stimulus that may affect the motion of the FV. The stimulus sensed and assessed by the FV as requiring a reaction is called the effective stimulus. In general, it takes a reaction time for the FV to respond accordingly to an effective stimulus, and the shortest reaction time $\tau_n^{\min}(t)$ (SRT) usually occurs in emergencies. While for a non-emergency stimulus, the FV's reaction time $\tau_n(t)$ is usually greater than $\tau_n^{\min}(t)$, and $\tau_n(t) \geq \tau_n^{\min}(t)$ is always true. For the FV, there is a maximum acceleration $a_n^{\max}(t)$ and a minimum deceleration $a_n^{\min}(t)$. While in motion, $a_n(t)$ is usually less than $a_n^{\max}(t)$ when accelerating and greater than $a_n^{\min}(t)$ when decelerating, $\left|a_n^{\max}(t)\right| < \left|a_n^{\min}(t)\right|$ is usually true, and $a_n^{\min}(t) \leq a_n(t) \leq a_n^{\max}(t)$ is always true. In general, the uniform motion is preferred for vehicles, but it is often difficult to keep $a_n$ at zero all the time even if conditions permit. These are the commonalities. All the parameters above as well as the bodywork length and the preferred speed of the uniform motion may be different for different vehicles, situations, environments, etc., which is where the heterogeneity lies. The complex phenomena exhibited by traffic flow are based on these facts.

*3.2.1. Effective stimulus and its composition*

The occurrence of an effective stimulus is the result of a comprehensive assessment of the stimulus composition. According to Table 1, in SMS. 1, the composition of the effective stimulus to decelerate the FV usually contains a decreasing spacing whose margin is not very large; while in SMS. 2, it usually contains the spacing itself or a negative rate of change of spacing. If the CDD is zero, then SMS. 2 does not exist. Therefore, if the PV decelerates continuously at a deceleration that reaches or approaches the minimum deceleration, or moves at a low or zero speed, then the effective stimulus to decelerate the FV should occur in SMS. 1. Regardless of the spacing margin stage, the composition of the effective stimulus to stop the deceleration of the FV usually contains a nonnegative rate of change of the spacing. And the composition of the effective stimulus to accelerate the FV usually contains a spacing greater than the APCCD. These are the commonalities.

It is worth emphasizing that the complete composition of the effective stimulus cannot be determined due to the heterogeneity, and hence the elements of the compositions identified above are usually the necessary conditions for the occurrence of the effective stimuli. These elements are closer to empirical results when the SRT and the minimum deceleration take the values that people are accustomed to rather than those determined by the physiology and the mechanical system.

*3.2.2. Vehicle reaction time and its composition*

After an effective stimulus is sensed and assessed by a driver or an intelligent driving system, a decision is made and driving operation is changed, and the motion of a vehicle usually responds with a very short delay. The total time from when an effective stimulus is sensed to when the motion determined by a decision takes place is the reaction time defined in this paper, and its composition is illustrated in Fig. 2. Therefore, the SRT of a vehicle may vary under different effective stimuli and decisions. For example, the SRT for throttling back is usually shorter than that for throttling back followed by applying brakes.

During the reaction time, the motion of a vehicle is related to the motion state when an effective stimulus is sensed, as well as the change in driving operation. If a new effective stimulus appears during this period, the changes in the driving operation may overlap. Therefore, the SRT of a vehicle may also vary under different motion states and driving operation. For example, to a new emergency stimulus, the SRT is the remainder of the initial SRT when in the reaction stage of the ERBP, and equals zero during the braking stage of the ERBP.

*3.3. Composition of the traffic flow property: inevitability*

In traffic flow, the decrease of the vehicle speed is caused by disturbances or certain traffic flow properties. Disturbances refer to lane changes, low-speed driving, ramps, slopes, bends, etc. that affect speed. When they appear, the FV may not be able to continuously replicate the spatiotemporal trajectory of the PV. That is, disturbances may propagate or grow in traffic flow, which is related to an element of the composition of the traffic flow property, i.e., inevitability.

*Outline of proof of inevitability:* According to Eqs. (2), (7) and (12),

$$\begin{aligned}M_n\left(t_n^e\right)+D_n^D\left(t_n^e\right) &= \int_{t_{n-1}^s}^{t_n^e} v_{n-1}(\tau)d\tau - L_{n-1} - D_n^C\left(t_n^e\right) \\ &= \min\left\{\int_{t_{n-1}^s}^{u} v_{n-1}(\tau)d\tau - \int_{t_n^e}^{u} v_n^{d_{\min}}(\tau)d\tau\right\} - L_{n-1}, \quad u \in I_n^{\min}\left(t_n^e\right).\end{aligned} \quad (16)$$

If the FV continuously replicates the spatiotemporal trajectory of the PV, then $v_n^{d_{\min}}\left(t_n^e\right)=v_{n-1}\left(t_{n-1}^s\right)$, $a_n\left(t_n^e\right)=a_{n-1}\left(t_{n-1}^s\right)$, and $h_n=t_n^e-t_{n-1}^s$ remains constant. It is clear that as $\int_{t_{n-1}^s}^{t_n^e} v_{n-1}(\tau)d\tau$ decreases or $D_n^C\left(t_n^e\right)$ increases, $M_n\left(t_n^e\right)+D_n^D\left(t_n^e\right)$ may be less than zero. And if this condition is satisfied, according to Theorem 4, the replication should terminate. If, further, the FV is to safely pass through where the front edge of the PV has passed, then it must and can only change its motion to increase $h_n$ or decrease $v_n^{d_{\min}}\left(t_n^e\right)$ or $a_n\left(t_n^e\right)$. According to Theorem 3, the inevitability of the disturbance propagation has been proved. If a disturbance propagates faster than it dissipates (i.e., the one at the forefront of the disturbed vehicles reverts to its original speed), then the section of traffic flow affected by the disturbance will expand, which is one form of the disturbance growth. Note that $M_n(t_n^e) + D_n^D(t_n^e)$ may still decrease when the time headway increases (see MS. 3, 6, and 12 in Table 2), so it may still decrease to be less than zero when the FV reaches the minimum speed that the PV has reached. In this case, the speed of the FV needs to continue to decrease, which is another form of the disturbance growth. Thus, the inevitability of the disturbance growth is proved.

*Note:* In reality, the motion of manned vehicles is more in the less stringent constraint space in Theorem 4, and drivers and vehicles are heterogeneous, which leads to probabilistic features in the propagation or growth of disturbances.

The composition of the traffic flow property comprises the commonality and heterogeneity of the motion of consecutive vehicles, and the inevitability of the disturbance propagation or growth. Under different conditions, these three elements determine specific changes and specific features in traffic flow phenomena, and thus determine different traffic flow properties.

*3.4. Inherent regularity of traffic flow*

According to Eqs. (12) and (16), the decrease of $M_n$ or $M_n + D_n^D$ is due to the decrease in the speed of the PV, or the decrease of the time headway, or the increase of the motion distance of the ERBP of the FV, while the increase of $M_n$ or $M_n + D_n^D$ is just the opposite. According to Theorems 1 and 3, the motion distance of the ERBP starting at a certain time is related to the motion state of the FV at the time, while the time headway at that time is related to the motion of the FV before the time. Therefore, when the FV transitions from one motion state to another, the change of the motion distance of the ERBP is fixed, that is, independent of the process, while the change of the time headway is not fixed, that is, related to the process. According to Theorem 4,

(1) as $v_{n-1}$ decreases, the potential minimum of $h_n$ increases;

(2) as $M_n + D_n^D$ decreases, the potential for $h_n$ to decrease decreases and the potential for $h_n$ to increase increases;

(3) as $D_n^D$ decreases, the potential minimum of $M_n$ increases.

In traffic flow, the FV is to change the motion state and the time headway according to the motion of the PV to maintain an appropriate $M_n$.

## 4. Brief reviews and conclusions

*4.1. Instability of traffic flow*

Traffic flow instability is one of the most important issues that has attracted widespread attention and academic interest among researchers, and a large number of studies have been proposed. The instability is considered to be related to: the reaction time greater than the anticipation time for the arrival of disturbance waves [16]; the time lag and delay caused by finite accelerations of vehicles and finite reaction times of drivers [17]; the lane changes that cause discharge rates to drop at bottlenecks [18]; the average vehicle density exceeding a critical value [19]; the congestion leading to the increased variance of density distribution [20]; the discontinuous character of the over-acceleration probability [21]; the combined effect of the jamming rate, jam lifetime, and jam size [22]; the travel speed exceeding a certain threshold [23]; the trucks mixed into traffic flow and the different types of car-following combinations [24, 25]; etc. These conclusions are drawn from different aspects.

The instability of traffic flow refers to the propagation and growth of disturbances. According to the composition of the traffic flow property, in terms of the inevitability, the instability is affected by the combination of factors such as the time headway, the travel speed, the bodywork length, the SRT, and the upper and lower limits of the speed change rate. Furthermore, the inevitability comes into play if certain conditions are met, and it is the commonality and heterogeneity that contribute to these conditions.

*4.2. Phase of traffic flow*

The introduction, development and debate of the phase theory have been around for a long time. Based on empirical traffic data, Kerner proposed that traffic flow has three phases with a fixed sequence of phase transitions (SPTs) [4], and criticized the generally accepted fundamentals and methodologies of the traffic and transportation theory for being inconsistent with empirical features [42]. Schönhof and Helbing pointed out that the classification of traffic states in Kerner's theory is not well-defined, still qualitative, and not general [8]. They believed that the three phases could not appropriately reflect the complexity of traffic phenomena, so they distinguished five states (two major categories) of the congested traffic based on the instability diagram and dynamic capacity, and gave quantitative phase diagrams under negligible and large perturbations, whose SPT differed from Kerner's theory [7]. So far, the study of the traffic flow phase has not actually been conclusive, but many important phenomena and properties have been identified, such as the pinch effect [4], the boomerang effect [43], and the nucleation nature of phase transitions [21].

In fact, the results given by previous studies are the partial cases of the traffic flow phase manifested in empirical results or simulations. Based on the composition of the traffic flow phenomenon and the stability of traffic flow, a definite theoretical basis for the division of the traffic flow phase can be given, and the complete phase will be revealed. The classification and identification of the phase can facilitate the determination of the traffic state and its changing trend, so related studies may be carried out in the future.

*4.3. Empirical datasets of traffic flow*

There is a consensus that empirical datasets are crucial to traffic flow studies. The existing datasets come from two sources: recordings of real vehicle trajectories and human driver data, and conducting experiments. The research of traffic flow has been greatly promoted by empirical data, a typical example of which is the trajectories provided by the Next Generation SIMulation (NGSIM) program [44]. Nonetheless, the accuracy of NGSIM data has also been questioned by researchers. Coifman et al. point out that large errors in NGSIM data have long been undetected by the research community, and even many experts lack first-hand experience in spotting such errors, which illustrates the scarcity of microscopic empirical data [45]. As early as 1999, Daganzo et al. pointed out that "no empirical studies to date describe the complete evolution of a disturbance through its entire life" [46]. Later, the empirical data from experiments contributed to the study of the traffic flow instability, but researchers argued that such data must take into account that drivers may have been unintentionally influenced by the experimental setup [7]. Until 2020, Li et al. wrote that traffic flow research is still eager for more accurate, higher quality, broader spatiotemporal scope, and more sourced (trajectories better integrated with human drivers' physiological and psychological measurement data, etc.) datasets [47]. In fact, the scarcity of empirical data has been a major bottleneck in the development of the traffic flow theory relying on empirical studies.

There has never been a single approach to scientific research. This paper has demonstrated that the fundamental theorems of traffic flow exist, and a basic theoretical framework can be obtained based on these theorems, on which the issues that have long puzzled researchers can be conclusive. The theoretical study and the empirical study should promote each other. In the future, empirical data can be used to bridge the gap from the fundamental theorems to applications, and its collection will be more targeted.

## Acknowledgment

This research was supported by the Fundamental Research Funds for the Central Universities, China (Grant No. 2021JKF421).

# Figures

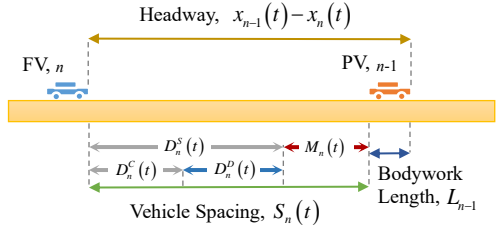

**Fig. 1. Illustration of two consecutive vehicles.**

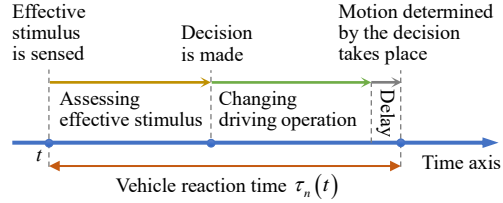

**Fig. 2. Composition of the vehicle reaction time.**

# Tables

**Table 1**
**3 stages of the spacing margin between two consecutive vehicles.**

| | | |
|---|---|---|
| $-D_n^D(t) \leq 0 \leq M_n(t)$ | SMS. 1 | If $M_n(t)$ is large, the FV has no need to react to the motion of the PV. The closer $M_n(t)$ is to 0, the more the FV needs to react *a priori*, and, further, the closer $-D_n^D(t)$ is to 0, the closer the reaction needs to be to the ERBP. |
| $-D_n^D(t) \leq M_n(t) < 0$ | SMS. 2 | Reaction of the FV to the motion of the PV is *a posteriori*, and the closer $M_n(t)$ is to $-D_n^D(t)$, the closer the reaction needs to be to the ERBP. |
| $M_n(t) < -D_n^D(t)$ | SMS. 3 | Inevitable collisions occur without lane changes. |

**Table 2**
**27 motion stages of two consecutive vehicles, where D, RC, and I are abbreviations for decreases, remains constant, and increases, respectively.**

| $M_n$ | | D | | | RC | | | I | | |
|---|---|---|---|---|---|---|---|---|---|---|
| $D_n^D$ | | D | RC | I | D | RC | I | D | RC | I |
| | D | MS. 1 | MS. 4 | MS. 7 | MS. 10 | MS. 13 | MS. 16 | MS. 19 | MS. 22 | MS. 25 |
| $h_n$ | RC | MS. 2 | MS. 5 | MS. 8 | MS. 11 | MS. 14 | MS. 17 | MS. 20 | MS. 23 | MS. 26 |
| | I | MS. 3 | MS. 6 | MS. 9 | MS. 12 | MS. 15 | MS. 18 | MS. 21 | MS. 24 | MS. 27 |